# Chip-scale terahertz frequency combs through integrated intersubband polariton bleaching

Francesco P. Mezzapesa,[1] Leonardo Viti,[1] Lianhe Li,[3] Valentino Pistore,[2*] Sukhdeep Dhillon,[2] A. Giles Davies,[3] Edmund Linfield,[3] and Miriam S. Vitiello[1]

[1]NEST, CNR - Istituto Nanoscienze and Scuola Normale Superiore, Piazza San Silvestro 12, 56127, Pisa, Italy
[2]Laboratoire de Physique de l'Ecole Normale Supérieure, Université PSL, CNRS, Sorbonne Université, Université de Paris, F-75005 Paris, France,
[3]School of Electronic and Electrical Engineering, University of Leeds, Leeds LS2 9JT, UK

* present address: NEST, CNR - Istituto Nanoscienze and Scuola Normale Superiore, Piazza San Silvestro 12, 56127, Pisa, Italy

**Abstract**

**Quantum cascade lasers (QCLs) represent a fascinating accomplishment of quantum engineering and enable the direct generation of terahertz (THz) frequency radiation from an electrically-biased semiconductor heterostructure. Their large spectral bandwidth, high output powers and quantum-limited linewidths have facilitated the realization of THz pulses by active mode-locking and passive generation of optical frequency combs (FCs) through intracavity four-wave-mixing, albeit over a restricted operational regime. Here, we conceive an integrated architecture for the generation of high power (10 mW) THz FCs comprising an ultrafast THz polaritonic reflector, exploiting intersubband cavity polaritons, and a broad bandwidth (2.3–3.8 THz) heterogeneous THz QCL. By tuning the group delay dispersion in an integrated geometry, through the exploitation of light induced bleaching of the intersubband-based THz polaritons, we demonstrate spectral reshaping of the QCL emission and stable FC operation over an operational dynamic range of up to 38%, characterized by a single and narrow (down to 700 Hz) intermode beatnote. Our concept provides design guidelines for a new generation of compact, cost-effective, electrically driven chip-scale FC sources based on ultrafast polariton dynamics, paving the way towards the generation of mode locked THz micro-lasers that will strongly impact a broad range of applications in ultrafast sciences, data storage, high-speed communication and spectroscopy.**

**INTRODUCTION**

The generation of stable frequency comb (FC) synthesizers with large optical powers per comb tooth [1], at terahertz (THz) frequencies (wavelength 300–30 μm), is fundamental for the investigation of light-matter interaction phenomena at the nanoscale, for quantum metrology [2], for communications [3], and for multiplexed analysis of gas samples requiring narrow-linewidth and a tight control of frequency jitter [4]. The most common technique to generate an FC in a solid-state laser is through mode-locking [5, 6]: the longitudinal modes of the laser cavity are locked in phase by means of an external (active) or internal (self- or passive) modulation mechanism, giving rise to a train of equidistant and intense pulses with a repetition rate equal to the inverse cavity round-trip time. Although mode-locked lasers have been widely demonstrated in the visible and near-infrared frequency ranges [5, 7], engineering passively mode-locked lasers in a compact and miniaturized architecture, across the THz frequency region of the electromagnetic spectrum, remains elusive.

Quantum cascade lasers (QCLs) have become, in the last decade, the most prominent electrically driven source of THz radiation, owing to their inherently high quantum efficiency [8, 9], compactness and spectral purity [10]. There is a fundamental obstacle preventing mode-locking with passive generation of ultra-short pulses in the semiconductor gain medium of a THz QCL: owing to the intersubband (ISB) architecture, the carrier relaxation is extremely fast (5-10 ps) [11]. As such, the gain recovery time is shorter than the cavity round-trip time (~70 ps for a 3 mm cavity) [12], complicating the generation of stable "ultrafast" laser pulses.

However QCLs, with specially designed heterogeneous [13, 14, 15, 16] or homogeneous [17] active regions, can perform as chip-scale THz FCs, characterized, in the frequency domain, by a set of equidistant spectral lines, which share a well-defined and stable phase relationship between one another. This is enabled by the large third-order $\chi^{(3)}$ Kerr nonlinearity of the active medium, which gives rise to the interaction between adjacent modes via four wave mixing (FWM) [18, 19] that generally results in a frequency modulated FC i.e. a quasi CW output. In a THz QCL the behavior is, in reality, more complex where both frequency and amplitude modulation are generally present [20], and act simultaneously.

One of the major drawbacks of THz QCL FCs is that the bias dependent group delay dispersion (GDD) usually compromises phase-locking of the laser modes over most of the laser operational range, therefore requiring proper dispersion compensation strategies. Successful approaches, commonly adopted in homogeneous frequency combs, include waveguide re-shaping [15], or integration with biased external elements [21]. Coupling with an unbiased external gold mirror has been conversely adopted to modify the reflectivity of heterogeneous FCs [22], providing dispersion compensation, although only over a limited portion of the QCL operational bandwidth. Dispersion is indeed very complex in heterogeneous THz QCLs and a simple metallic mirror cannot compensate the GDD over the entire laser bandwidth.

Semiconductor mirrors, relying on inherently fast intersubband polariton dynamics, could be in principle ideal, in this respect. They simultaneously provide an easier self-integration in the QCL cavity, while still maintaining a significant flexibility in terms of design and spectral bandwidth. Furthermore, recently, we have demonstrated that by strong coupling of ISB transitions [23, 24] of semiconductor quantum wells to the photonic mode of a metallic cavity, we can custom tailor the population and polarisation dynamics of ISB cavity polaritons in the saturation regime [25, 26]. This results in efficient solid-state mirrors operating in the 2–3 THz range [25, 26] that can provide a strong modulation of the optical response on sub-cycle timescales, accompanied by a recovery time ~ 3.3 ps.

In the present work, we conceive an all-solid-state architecture in which the polaritonic mirror is integrated with a set of broadband heterogeneous THz QCLs, having dissimilar dimensions. We first match the polaritonic doublet with the heterogeneous QCL bandwidth. Then, by exploiting the complex reflectivity change associated to the light-induced bleaching of intersubband polaritons, we tune the GDD, achieving compensation over a bandwidth and a current range much larger than that reached so far in any THz QCL FC. The fabricated devices show stable operation as optical FCs over an operational range > 35%, significantly larger than that spontaneously achievable (12%–15%) from the corresponding bare QCL [14], providing ~10 mW of continuous-wave (CW) output power, ultra-narrow (760 Hz) intermode beatnote linewidths (LWs), and > 80 equally spaced optical modes covering a maximum bandwidth of 0.85 THz in the comb regime and of 1.5 THz in the dispersion dominated regime. Furthermore, we show that the integrated polaritonic mirror can reshape the QCL emission spectrum leading to a significant mode proliferation at half of the laser operational range (i.e. current density $J = 1.43\ J_{th}$, where $J_{th}$ is the threshold current density), with a corresponding narrow intermode beatnote. This is the signature of phase locking of the laser modes in a regime where the heterogeneous nature of the gain media spontaneously entangles the dispersion dynamics.

**EXPERIMENTAL RESULTS**

**Samples description, experimental setups and simulations**

The polaritonic mirror is based on a semiconductor multi-quantum-well (MQW) heterostructure, resonant at the ISB transition frequency $\nu_{ISB} = 2.7$ THz [25, 26]. The MQW stack is embedded in a ~2 μm cavity, consisting of a back Au reflector and a top Au grating with period 16 μm. The latter provides the required optical coupling to the MQW. When radiation polarized orthogonal to the grating lines (*p*-polarization) impinges at normal incidence, a fringe electric field is localized within the MQW heterostructure at the metallic edges of the grating. This field distribution satisfies the ISB transition selection rule in the near-field via the non-vanishing component $E_z$ along the MQW growth direction [26].

The reflectance of the polaritonic mirror, measured in vacuum at low temperature (6 K) for normal incidence (Fig. 1a), reveals a characteristic polaritonic doublet [27] separated by a Rabi frequency of 0.18 THz. We exploit the ultrastrong light-matter coupling of the designed intersubband transition (black curve Fig.1a) to the near-field of the metallic grating; the strong field enhancement induced by the reduced mode volume significantly enhances the optical absorption when compared to the bare electronic transition, therefore leading to a reduced saturation power in the two-state polaritonic system. By pumping the polaritonic mirror with a CW, 2.75 THz laser, we observe a visible spectral change of the reflectivity at intensities larger than 7.8 Wcm$^{-2}$, reflected in the bleaching of the upper polaritonic mode [26]. This results in a visible reflectivity change in the 2 – 3 THz spectral window, as shown in Fig. 1a, where three prototypical reflectivity curves, corresponding to driving currents of the incident QCL of 520 mA, 580 mA and 632 mA, are shown.

Such an effect can be innovatively exploited to tailor the intracavity dynamics of broadband QCLs by modifying their cavity dispersion. To this end, we use the polaritonic mirror as the back mirror of a set of free-running THz QCL frequency combs (Fig. 1b), exploiting two different heterogeneous active regions (ARs), which allow broadband operation over a bandwidth (2.3 THz – 3.8 THz) that matches, on its low frequency side, the spectral position of the polaritonic doublet (Fig. 1a). This ensures that the reflectivity spectrum of the polaritonic doublet partially overlaps with the gain profile of the QCLs used (Fig. 1c). The variation of the peak value of the gain profile is used in the simulations to reproduce the increase of the dispersion for higher currents, which occurs even though the maximum gain is clamped to the total losses [28].

The two GaAs/AlGaAs QCL heterostructures each comprise three active modules, exploiting alternating photon- and longitudinal optical (LO) phonon-assisted interminiband transitions [29], individually designed to operate at a different central frequency (2.5 THz, 3.0 THz, 3.5 THz). The two ARs differ in the doping concentration ($n_d = 3.2 \times 10^{16}$ cm$^{-3}$ for laser A, and $n_d = 4.0 \times 10^{16}$ cm$^{-3}$ for laser B), which is optimized to achieve a flat gain bandwidth (Figure 1c for laser A) and uniform power output across the whole spectrum, and to obtain two equivalently high dynamic ranges of current density ($J_{max}/J_{th} = 2.9$). Both lasers are fabricated in a metal-metal waveguide configuration with a set of nickel side absorbers [14] that have the specific purpose of inhibiting lasing from higher order lateral modes. Laser A is 85 μm wide and 2.9 mm long, laser B is 50 μm wide and 2.3 mm long. The comparison between the light-current density-voltage (LJV) characteristics of laser A [14] and B (Figure 1d) reveals a doping-dependent threshold current density increase, varying from 150 Acm$^{-2}$ in sample A, to 175 Acm$^{-2}$ in sample B.

The polaritonic mirror is mounted, together with the QCL, on the copper cold-unit of a helium-flow cryostat and tightly coupled to the QCL back-facet at a distance $d_c \sim 50 \pm 2$ μm, determined by a fixed spacer. This defines a Gires-Tournois interferometer (GTI) created by the external cavity between the QCL back-facet and the surface of the polaritonic sample [30, 22], which is intentionally pre-defined to create a chromatic dispersion opposite to that arising in the QCL cavity within the frequency range of interest. The polaritonic grating is oriented in a direction orthogonal to the QCL light polarization direction (*p*-polarization) to excite the MQW intersubband transitions.

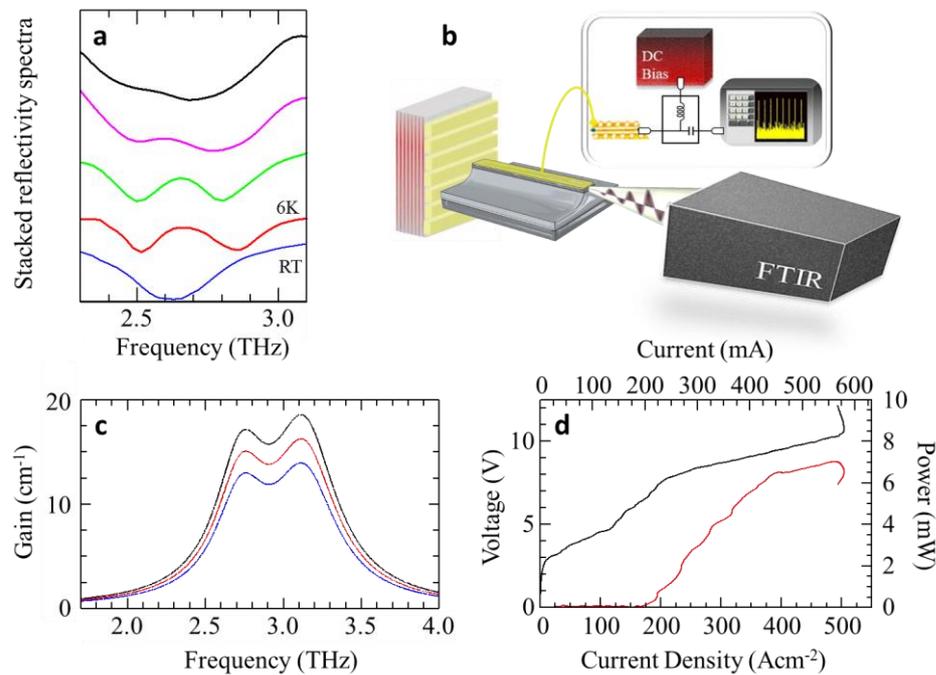

**Figure 1**. (a) Reflectivity spectra of the polaritonic structure measured at T = 300 K (blue curve), and at T= 6 K while pumping on it a continuous-wave QCL (laser A) driven at $I$ = 0 mA (red curve), $I$ = 520 mA (green curve), $I$ = 580 mA (pink curve), $I$ = 632 mA (black curve). (b) Schematic diagram showing the experimental arrangement. (c) Gain profile of laser A, extrapolated from the corresponding experimental emission spectra collected at driving currents $I_A$ = 520 mA; $I_A$ = 580 mA; $I_A$ = 632 mA (from bottom to top). (d) LIV characteristics of laser B. Analogous plots for laser A are reported in ref. [14].

To show the effect of dispersion compensation from the polariton mirror, we perform numerical simulations of the group delay dispersion (GDD). The dispersion profile includes the contributions from the material and gain of the QCL, as well as that of the GTI. The first two terms are computed using a Drude-Lorentz model for the frequency dependent refractive index of the material; by using the Kramers-Kronig relations [28], we then add the correction due to the QCL gain, the latter retrieved from the experimental emission spectra. The waveguide dispersion contribution is negligible with respect to the other terms and is therefore not considered here. The dispersion profile introduced by the GTI (Figs. 2a-c) is then obtained following the procedure described in Ref. [30] for a non-ideal GTI, where the frequency dependent reflectivity of the polaritonic reflector varies with the driving current of the laser according to the FTIR reflectivity spectra of Figure 1a. Since the radiation emitted from the QCL facet is not entirely coupled back into the waveguide, it is also necessary to account for the optical feedback from the polaritonic mirror. This is extrapolated from numerical simulations using a finite element method (Comsol Multiphysics). The simulated structure includes the end of the QCL waveguide and the polaritonic mirror placed at a 50 μm distance from the laser facet, surrounded by vacuum. Appropriate absorbing boundary conditions are set at the external boundaries of the simulation domain. THz radiation is injected into the QCL waveguide (into the end opposite to the GTI) and is reflected back into the QCL waveguide by the external mirror. This allows one to obtain the amplitude and phase of the scattering parameter and estimate the optical feedback.

Figures 2a-c show the dispersion profiles due to the gain and the material (in black), the GTI with the polaritonic mirror (in blue), and the sum of all the contributions (in red), calculated at different driving currents. The reduction of the overall dispersion was evaluated by computing the average of the absolute value of the dispersion profiles in the lasing range (2.55 THz – 3.25 THz). Without the GTI, the average dispersion due to the gain and the material increases from $4.99\times10^5$ fs² at $I_A$=520mA to $5.58\times10^5$ fs² at $I_A$=580mA, and then to $6.18\times10^5$ fs² at $I_A$=632mA. By employing a GTI, the average dispersion is $3.03\times10^5$ fs² at $I_A$=520mA, $2.27\times10^5$ fs² at $I_A$=580mA, and $4.11\times10^5$ fs² at $I_A$=632mA. The driving current where the lowest dispersion is achieved is ~580mA; that is also where the LW of the experimentally-measured beatnote approaches 700 Hz.

We then run a further set of simulations while driving laser A at higher currents. Figure 2d shows the simulation results at $I_A$ = 800 mA. As clearly visible from the plot, the polaritonic GTI is unable to compensate or even significantly reduce dispersion, except in a small spectral window (2.8-3.1THz), with a visible major GDD increase at frequencies <2.8 THz, meaning that no major effects on the QCL mode behavior is expected in such a biasing regime.

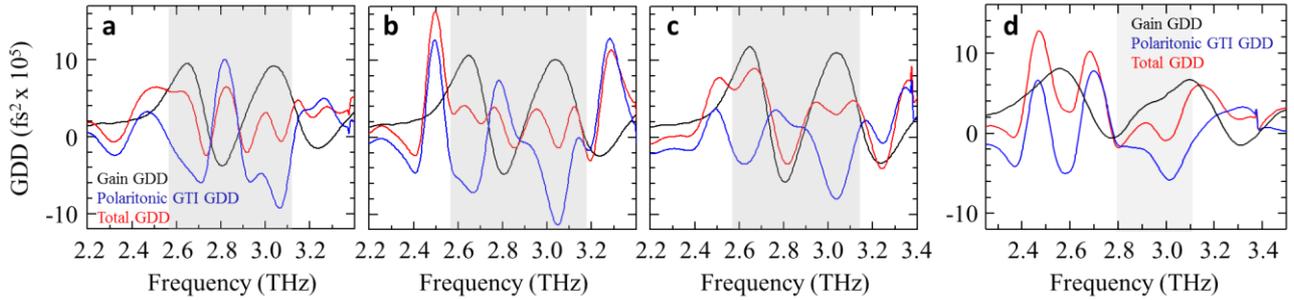

**Figure 2**. (a-d) Simulated group delay dispersions: GDD of the QCL gain (black curve), GDD of the polaritonic mirror (blue curve) and total GDD (red curve) of the resulting GTI calculated for a GTI length of 50 μm when the QCL is driven with a current $I_A$=520 mA (a), $I_A$=580 mA (b), $I_A$=632 mA (c) and $I_A$=800 mA (d). The shaded grey areas mark the spectral regime in which the total GDD remains lower than the QCL GDD, leading to GDD compensation (a-c) or to a negligible effect on the overall GDD (d).

**Results and discussion**

The QCLs are electrically driven in CW at a heat sink temperature $T_H$ = 15 K, over a bias-tee, which allows the free running electrical beatnote of the QCL combs to be monitored using an RF spectrum analyzer (*R&S FSW43*). The output radiation is aligned with an in-vacuum Fourier transform infrared spectrometer (FTIR, Bruker Vertex 80v), with a resolution of 0.075 cm$^{-1}$. Under this experimental configuration, we can simultaneously record the output THz spectrum of the sources and their RF power spectral density (PSD).

We first test the operation of the polaritonic mirror by coupling it to laser A [14], which naturally behaves like a comb in the current ranges $I_A$ = 430 mA – 536 mA and 540 mA – 543 mA [14]. The multimode THz spectrum consists of equidistant modes, which beat together, causing a modulation of the laser intensity at a frequency in the range 13.7 GHz – 14.0 GHz. We span the operation current of the laser A ($I_A$) over the whole dynamic current range ($\Delta I_A$ = 670 mA) while collecting the intermode beatnote and the corresponding FTIR emission spectra.

Figure 3a shows the intermode beatnote map of the coupled system, as a function of $I_A$. A single beatnote is observed in a continuous range between 430 mA and 543 mA, then again over a small current portion (568 mA – 575 mA), and finally from 582 mA to 660 mA.

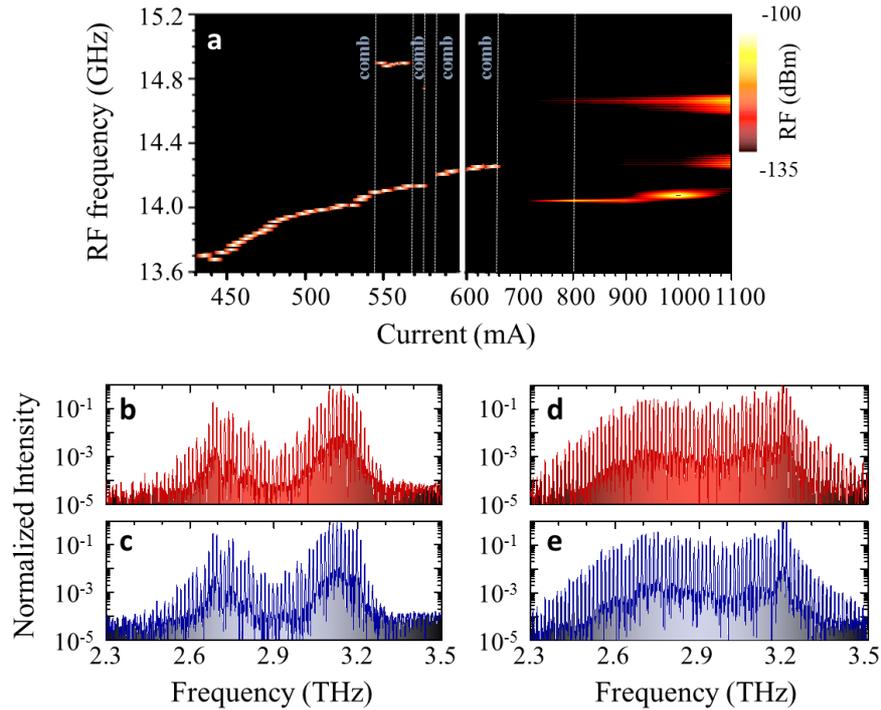

**Figure 3**: Laser A. Spectral characterization and intermode beatnote spectroscopy. **(a)** Electrical RF intermode beatnote map measured at 15 K, as a function of the QCL bias current for a 2.9-mm-long, 85-μm-wide laser bar coupled at a distance $d_c$ = 50 μm to a polaritonic saturable absorber mirror; **(b-e)** FTIR emission spectra collected at 15 K, while driving the QCL in CW at $I_A$ = 650 mA (b,c) and at $I_A$ = 800 mA (d,e) in the coupled QCL-polaritonic mirror system (b,d) and in the bare laser A (c,e), respectively.

Overall the single beatnote persists over 30% of the laser dynamic range, i.e. over a region twice as wide as that retrieved in the bare laser A; this is a signature that the dispersion compensation induced by the polaritonic reflector allows phase locking of the lasing modes over a wider operational regime. A dual beatnote is then observed in the current range 543 mA – 568 mA, i.e. over a current portion slightly larger than that retrieved on the bare laser [14]. Such a regime reflects the dual comb nature of our heterogeneous active core, with the two families of optical modes centered at ~3.1 THz and ~2.7 THz (Fig. 3b-3d) clearly visible at low driving currents. Finally, at currents larger than 660 mA we observe a broad beatnote, which is the signature of a lasing regime where the GDD is large enough to prevent the FWM from locking the lasing modes, in both frequency and phase, simultaneously.

The comparison between the FTIR spectra of the bare laser A (Fig. 3c, 3e) and those of the coupled laser system (Fig. 3b, 3d) do not reveal any visible change either in the regime where the coupled system shows a single beatnote (Figs. 3b-3c) and in the high current regime, dominated by dispersion (Figs. 3d-3e). Conversely, the comparison between the corresponding beatnotes (Fig. 4a) clearly shows the efficacy of the proposed approach. At $I_A$ = 650 mA, a single, 30 dBm intense RF signal is retrieved, while the bare laser A shows a very broad beatnote [14].

To confirm that the measured phenomena are correctly ascribed to the intersubband polaritonic grating, we repeated the same experimental procedure after turning the polaritonic absorber by 90°, so that the polarization of the incoming THz beam is unable to activate ISB transitions in the multi-quantum well grating structure (*s*-polarization). In this configuration, which basically corresponds to the case of a reflecting gold grating (the ISB absorption is strongly suppressed [26]), the beatnote spectrum remains unaltered with respect to that of the bare laser [14]. The analysis of the beatnote LW across the current dynamic range of laser A coupled to the polaritonic mirror in the *p*- and *s*- polarized cases is shown in Figure 4b. We evaluate the beatnote LW by fitting the acquired RF spectra with a Lorentzian distribution (in the case of single beatnote) or by determining the half-width-at-half-maximum (HWHM) of the power spectral density distribution (in the cases of broad beatnote). The coupling with the polaritonic mirror (*p*-polarization) efficiently reduces the intermode beatnote LW over the whole laser dynamic range, with respect to the case of a QCL coupled with a gold grating double metal cavity (*s*-polarization), which conversely does not induce any change on the intermode beatnote LW with respect to the case of the bare [22] laser. At currents $I_A$ < 543 mA (region I, Fig. 4b), the RF spectrum shows a single narrow beatnote for both *p* and *s* polarizations, meaning that laser A is behaving like a comb, regardless of the coupling configuration adopted. Such an effect is expected since, even if the ISB transition is not activated, the polaritonic grating is here behaving as a gold-like reflector which, when in the "on-resonance" GTI configuration [22] ($d_c$ ~ 50 μm), is expected to compensate the QCL GDD, although over a partial frequency window (2.7-3.1 THz), as we experimentally demonstrated on the same laser bar [22]. However, once coupling the QCL with the polaritonic mirror (*p*-polarization), the LW is significantly reduced by more than a factor of five, reaching a minimum of 760 Hz at $I_A$ = 560 mA (Figure 4c).

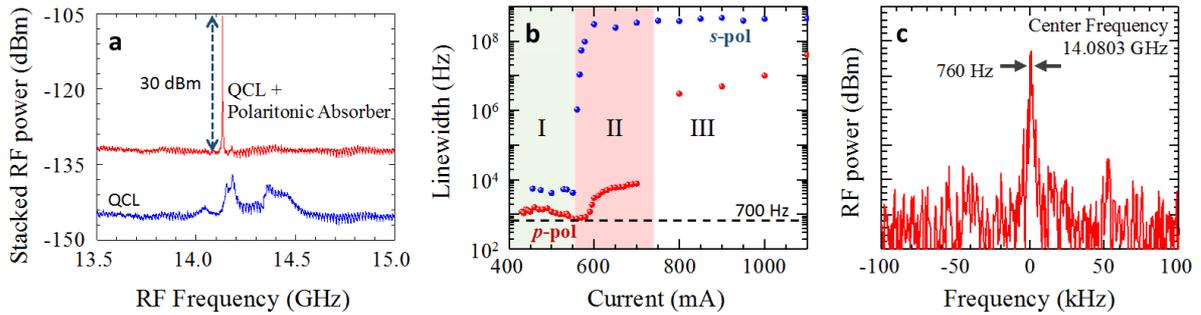

**Figure 4**. (a) Stacked electrical intermode beatnotes acquired at $T_H$ = 15 K, while driving the bare laser A and the same laser coupled with the polaritonic mirror at driving currents of 650 mA. The RF spectra have been shifted vertically by 15 dBm. (b) Intermode beatnote linewidth of laser A coupled with the polaritonic mirror under *p*-polarization (red) or *s*-polarization (blue), plotted as a function of the driving current. The blue curve is almost coincident with that retrieved on the bare laser A [22]. (c) Intermode beatnote measured while driving the coupled laser system at a current of 550 mA. The shaded areas, labeled as I, II, III identify different transport regimes in which the laser intracavity dynamics changes.

This result is explained by the simulation of Figure 2a, which shows that the reflectivity of the intersubband polaritonic mirror induces a reduction of the total GDD with respect to that of the bare laser A over a wider spectral window (2.55–3.15 THz), matching 80% of the spectral emission of the coupled system (Fig. 3b). The LW of the main beatnote (even when a double beatnote appears) remains < 850 Hz until a driving current of 582 mA. Above 582 mA neither the bare QCL nor the QCL coupled to the *s*-polarized mirror presents an individual narrow beatnote. This means that the group velocity dispersion is sufficiently strong to break the intermode coherence generated by the intracavity FWM process [14] and that the GTI comprising the gold grating structure even in the on-resonance condition is unable to compensate dispersion. The latter effect is in full agreement with previous experimental reports [22]. However, in the coupled system comprising the QCL and the polaritonic mirror in *p*-polarization, at 582 mA, the narrow individual beatnote regime persists, which is signature of the fact that the GDD is compensated by the modulation of the losses induced by the polaritonic mirror. This is reflected in the simulation results (Fig. 2b) which shows that, at 580 mA, the total GDD is drastically reduced and compensated over a spectral window extending from 2.55 THz to 3.2 THz, i.e. matching almost the whole emission bandwidth of the QCL (Fig. 2b). At driving currents in the range 582–660 mA (region II, Figure 4b), the beatnote LW slightly increases (900 Hz – 7 kHz) as a consequence of the predicted GDD increase (Fig. 2c), but remains single, confirming the prediction of the simulations (Fig. 2c), showing that the total GDD of the integrated QCL-polaritonic mirror system remains lower than the GDD of the bare laser over a slightly smaller spectral window 2.58 THz – 3.1 THz.

Finally, at currents larger than 660 mA (region III), the beatnote becomes wider. Interestingly, a visible reduction of the phase noise (by more than one decade) is observed. In contrast, the GTI comprising the QCL and the gold-grating double metal cavity (*s*-polarization) does not show any beatnote for any current value in the 550 mA- 1050 mA range (Fig. 4b, regions II and III).

To further confirm our claim that the GDD is sensibly compensated by the modulation of the losses induced by the polaritonic mirror, we repeated the same experiment by varying the distance between the polaritonic mirror and the QCL back facet under both polarizations (*p*-pol and *s*-pol). First, we slightly detuned (by 5 µm) the position of the mirror and collected the bias dependent intermode beatnote under the polarization state where intersubband absorption is not activated (*s*-pol). The evolution of the intermode beatnote linewidth remains practically unperturbed with respect to the bare laser case (see Fig. 4c and 5), confirming that the mirror is here behaving as a simple gold mirror in an on-resonance GTI configuration, in complete agreement with previous reports [22]. We then varied the distance of the polaritonic mirror up to 80 µm, therefore defining an off-resonance GTI [22]. We collect the beatnote map under both polarizations (Fig. 5). The results clearly show that no dispersion compensation is achieved neither in *p*-pol nor in *s*-pol configurations, meaning that the off-resonance GTI is ineffective to induce a visible dispersion compensation neither in the case of light-activated intersubband transitions, nor when the mirror is behaving as a simple gold grating [22].

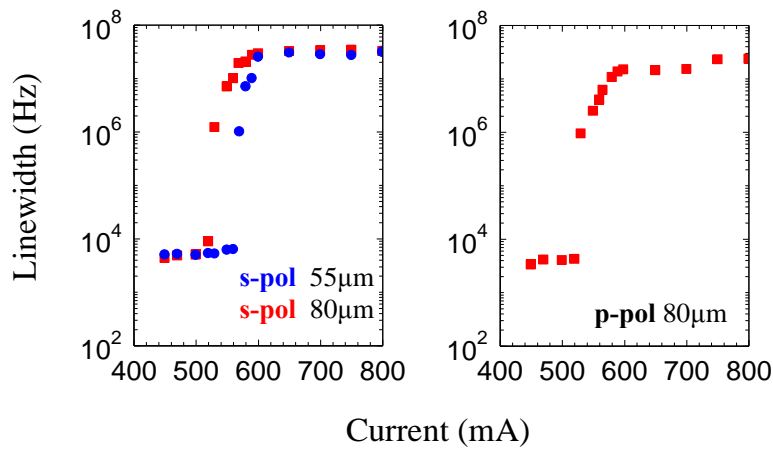

**Figure 5. (a)** Intermode beatnote linewidth plotted as a function of the driving current of laser A coupled with the polaritonic mirror under *s*-polarization and while placing the polaritonic mirror at 55 µm distance or at 80 µm distance from the back QCL facet. **(b)** Intermode beatnote linewidth plotted as a function of the driving current of laser A coupled with the polaritonic mirror under *p*-polarization and while placing the polaritonic mirror at 80 µm distance from the back QCL facet.

We then coupled the polaritonic mirror with laser B. We first characterize the spectral behavior of the bare laser. By spanning the driving current $I_B$ of the bare laser over the whole dynamic current range, i.e. from 200 mA (threshold current) to 490 mA (roll-off current), we find four distinct lasing regimes. The laser is initially single mode (emitting at 3.23 THz) for $I_B$ < 210 mA (Fig. 6a). Then, up to $I_B$ = 270 mA, only the higher-frequency stage of the heterogeneous active medium is above threshold, with modes separated by the cavity round-trip frequency $f_{rt}$ = 17.0 GHz (Figure 6b). In this regime, the RF spectrum shows a single and narrow beatnote at ~ 17 GHz; the intermodal frequency varies from 16.2 GHz at $I_B$ = 210 mA to 17.5 GHz at $I_B$ = 270 mA, with LW ~ 20 kHz (Figure 6e). At higher $I_B$ (between 270 mA and 320 mA), the lower-frequency AR module crosses threshold. In this range, the QCL operates in a harmonic state [17, 32, 33,34], where the spacing between adjacent modes is 2 $f_{rt}$ for the higher frequency AR module, whereas it is equal to 6 $f_{rt}$ for the lower frequency module (Figure 6c). This regime is driven by the interplay of the third-order population pulsation (PP) nonlinearity and the population grating (PG) induced in the cavity by the primary mode of the lower frequency AR [32], the intense mode at 2.73 THz in Figure 5c. Basically, this single-mode instability is the result of the standing-wave of the primary mode, which induces a spatial modulation of the population inversion (spatial hole burning), whose interaction with adjacent optical modes can suppress neighboring Fabry-Perot modes, while favoring sidebands that are some (or several) $f_{rt}$ apart. The third-order nonlineari1ty of the active medium can simultaneously give rise to a temporal modulation of the 10population inversion. As a result of the PP and PG combination, the QCL shows a beatnote, which can be the signature of an amplitude modulation (AM) of the THz wave in the time domain if the PP effect is dominant or of a frequency modulation (FM) when the PG is dominant. We argue that in our device both mechanisms are simultaneously present [28].

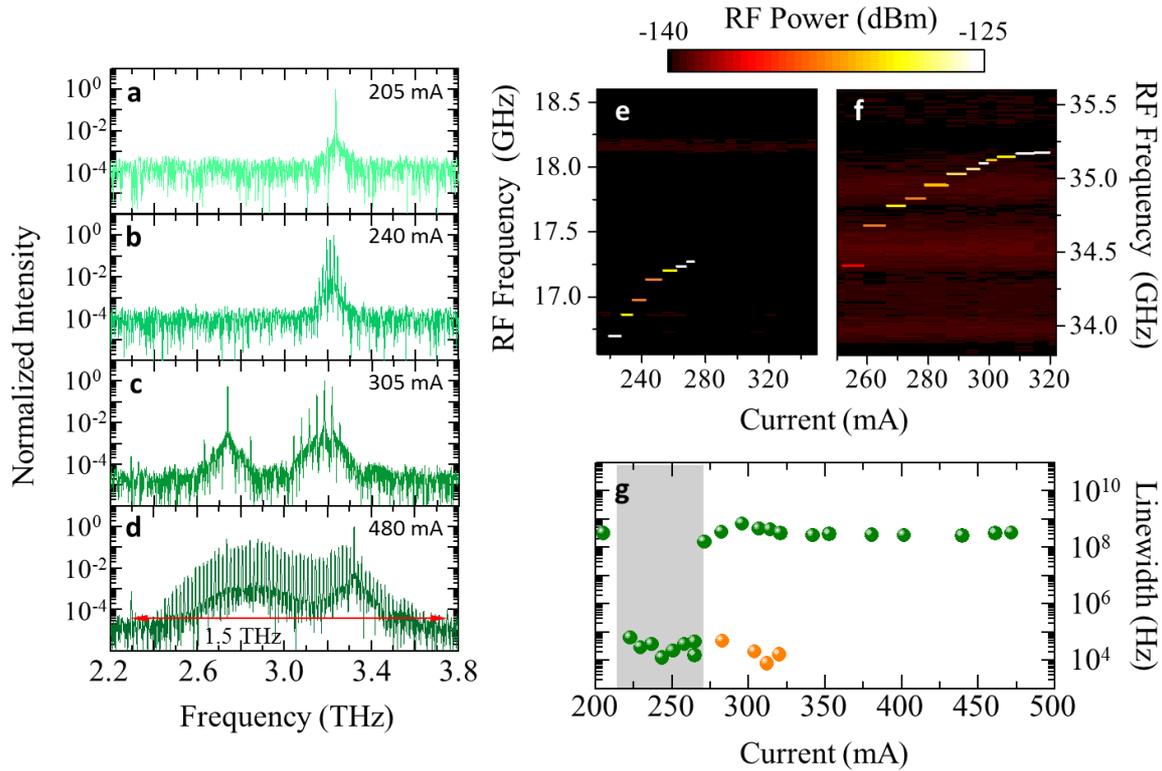

**Figure 6**. Spectral characterization and intermode beatnote spectroscopy of laser B. **(a-d)** FTIR emission spectra of the bare laser B, recorded at $T_H = 15$ K and at driving currents of 210 mA (a), 240 mA (b), 305 mA (c), and 480 mA (d). **(e)** Intermode RF spectrum recorded in the range 210 mA < $I_B$ < 335 mA; **(f)** Beatnote map recorded when laser B is operated in the harmonic comb regime (260 mA < $I_B$ < 325 mA), showing a visible peak at ~35 GHz. For a small current range 260 mA < $I_B$ < 270 mA, the beatnotes at 17 GHz and 34 GHz coexist. **(g)** Intermode beatnote linewidths plotted as a function of the driving current (green dots indicate the standard frequency comb, orange dots indicate the harmonic state). The grey shaded area shows the region in which the device behaves as a frequency comb, showing a single, narrow beatnote.

The RF spectrum in this regime is characterized by a narrow beatnote at a frequency $2f_{rt}$. Our experimental arrangement does not allow us to determine whether there is another beatnote at $6f_{rt}$, and no beatnote is observed at 17 GHz. The RF spectrum is shown in Figure 6f, with a stable and narrow beatnote with LW ~8 kHz, demonstrating the coherence of the adjacent modes in the high frequency portion of the spectrum.

For $I_B$ > 320 mA, all the modules of the heterogeneous AR are activated and a broad (1.5 THz, with a continuous sequence of equally spaced optical modes covering 1.38 THz) and dense spectrum is observed up to $I_B$ = 480 mA (Figure 6d), corresponding to an emitted optical power of 7.2 mW. In this regime, dominated by cavity dispersion, a broad beatnote is retrieved, which is the signature of the lack of phase coherence between the lasing modes. The plot of the $f_{rt}$ beatnote linewidths (Fig.6g) shows that laser B is behaving as a stable frequency comb over a dynamic current range ($\Delta I_B$) = 15%.

We then coupled laser B with the polaritonic mirror, at a distance of 50 μm, both in the conventional orientation (*p*-polarization) and in the *s*-polarization configuration where the intersubband transition in the polaritonic region cannot be activated. The comparison between the intermode beatnote maps (Figs. 7a, 7d) shows that, in the *s*-polarized case, the behavior of the QCL approximately matches that of the bare laser *B*, with a single beatnote persisting in the 220 mA–275 mA range. The beatnote LW values are also comparable (~ 10 kHz) with those found in the bare laser B (Fig. 6g). We find that, similarly to what happens in the case of the bare laser B, when $I_B$ is driven above 270 mA, the QCL shows 15 optically active modes separated by $2f_{rt}$ (Figure 7f). At currents larger than 270 mA, the laser does not behave as a comb over the remaining current dynamic range.

The picture drastically changes when the polaritonic mirror is integrated with the QCL, with the correct coupling geometry (*p*-polarization). The intermode beatnote map shows an individual narrow (LW~10 kHz) beatnote for $I_B$ < 270 mA, as shown in Figures 7c-7d. A second beatnote (at 17.6 GHz) appears in a very small current range (275 mA – 280 mA), just above the onset of laser action of the second active region. This is a signature of the persistence of a dual-comb operational regime, in which the two individual active region stacks are individually locked in phase with slightly detuned intermode frequencies, although not phase-locked together.

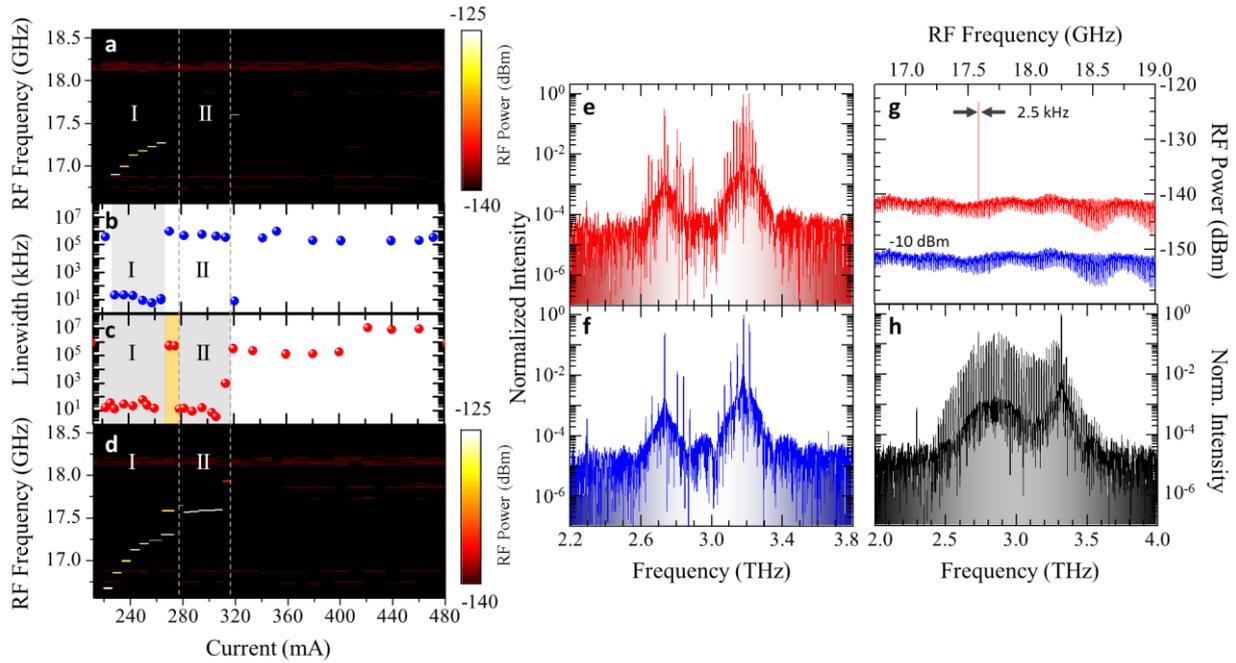

**Figure 7**. Spectral characterization and intermode beatnote spectroscopy of the integrated laser system. **(a)** RF beatnote map recorded as a function of the driving current when the polaritonic mirror is positioned at the QCL back-facet in *s*-polarization. **(b)** Beatnote *LW* for coupling under *s*-polarization. The grey shaded area indicates the regime of stable comb operation. **(c)** Beatnote *LW* for coupling under *p*-polarization; and **(d)** corresponding beatnote map. The comb operation is extended over the current range 270 mA – 320 mA. The transition from the lower FC current range (region *I*) to the higher FC current range (region *II*) is marked by the presence of two narrow beatnotes at 17.3 GHz and at 17.6 GHz (yellow shaded area, in panel c). **(e-f)** THz spectra recorded at $T_H = 15K$, while driving the QCL in CW with a current of 305 mA for *p*- (e) and *s*- (f) polarization states. **(g)** Stacked RF spectra measured while driving the QCL in CW with a current of 305 mA for the *p*- (red trace) and *s*- (blue trace) polarization states. **(h)** FTIR spectrum recorded in CW at $I_B = 480$ mA, while keeping $T_H = 15K$.

Then, for 280 mA < $I_B$ < 320 mA, the beatnote returns again to be single and narrow (2.5 kHz - 8 kHz) and a large number of lasing modes (40), spectrally spaced at $f_{rt}$, appear (Fig. 7e); therefore, the polaritonic mirror allows the laser to be driven from a harmonic comb state (Figs 6c) to a phase-stabilized frequency comb. Correspondingly, a visible intermode beatnote LW reduction down to 2.5 kHz at $I_B = 305$ mA is detected (Fig. 7g). This result is ascribed to the non-trivial response of the polaritonic mirror to the amplitude modulation of the harmonic comb output. The optical feedback from a simple reflector, such as the *s*-polarized grating or a gold mirror is insufficient to compensate the single-mode instability that drives the laser in the harmonic state. On the other hand, the strong compensation of the GDD induced by the *p*-polarized polaritonic mirror is capable of preventing the instability from taking place. In this regime, simulations predicts a visible reduction of the GDD (see Fig. S1b, Supplementary Information).

Finally at $I_B = 320$ mA, when the third stage of the AR (centered at 3.0 THz) reaches the lasing threshold, the single beatnote frequency abruptly shifts to 17.9 GHz, while still preserving its narrow nature.

For currents > 320 mA the coherence between the modes is lost and a broad beat note (LW > 100 MHz) is observed, in agreement to what predicted by simulations (see Supplementary Information). The corresponding laser spectrum (Figure 7h) shows 80 optically active modes covering a bandwidth of 1.25 THz (2.3 THz – 3.55 THz).

Remarkably, the integration with the polaritonic mirror leads to an overall increase of the comb dynamic range operation from 13% in the case of the bare laser to ~ 30%, in full agreement with that measured with laser A.

**Conclusions**

In conclusion, we demonstrate that by engineering an intersubband polariton saturable absorber reflector, with dynamics considerably faster than the gain recovery time of QCLs [27], and by coupling it with broadband THz QCLs, stable optical FCs are generated, characterized by narrow free running intermode beatnote LWs (700 Hz) and up to 10 mW of emitted optical power, spread over ≥ 40 teeth. We have shown that the polaritonic mirror behaves as a novel and efficient dispersion compensator of the complex gain profile of a heterogeneous QCL. Further, as the polariton mirror can simultaneously act as a saturable absorber mirror [27], our experimental results can potentially open a path towards passively THz mode-locked micro-cavity lasers in a monolithic single-chip design with wide implications for: metrology, where laser excitation can match the energy levels splitting of molecules and its pulsed nature can down-convert the spectrum to the RF domain; ultrafast communications, where THz frequency carriers are requested for high-bandwidth data transfer; and, THz quantum optics, where high-power pulses can drive molecular samples out of equilibrium.

**Methods**

**QCL Fabrication**

The QCL is processed in a double-metal configuration starting from Au-Au (400 nm/400 nm) wafer bonding via thermo-compression on a highly-doped GaAs carrier (or receptor) wafer. The bottom highly-doped GaAs contact layer is then exposed through a combination of mechanical thinning and selective wet-etching, after removing the $Al_{0.5}Ga_{0.5}As$ etch-stop layer. The 17-μm-thick active region is then defined by dry etching in an inductively-coupled plasma reactive ion etching (ICP-RIE) facility. Dry etching allows almost vertical sidewalls, which result in uniform current injection and reduced lateral optical scattering. A Cr(10 nm)/Au(150 nm) top contact is then defined by a combination of optical lithography (SUSS MicroTec MJB4), thermal evaporation and lift-off. The top contact is intentionally patterned narrower than the ridge top surface, allowing thin nickel side-absorbers (setbacks) to be deposited via optical lithography on the uncovered portion of the ridge surface, using a laser writer (MicroWriter ML3 Durham Magneto Optics) [14]. The side-absorbers are 5 μm wide and 5 nm thick for laser A, and 2 μm wide and 5 nm thick for laser B. Laser bars are finally cleaved, mounted on a copper bar with an indium-based thermally conductive adhesive paste, wire bonded with an ultrasound wedge bonding and connected to high frequency coplanar waveguides.

**Acknowledgments**

**Funding:** This work was partially supported by the European Research Council through the ERC grant 681379 (SPRINT), the European union FET open project ULTRAQCL (665158), and by the EPSRC (UK) programme 'HyperTerahertz (EP/P021859/1). EHL acknowledges support of the Royal Society and Wolfson Foundation.

**Competing interests:** The authors declare no competing financial interests.